\documentclass[twocolumn,showpacs,preprintnumbers,pra,amsmath,amssymb, superscriptaddress]{revtex4-1}

\usepackage{graphicx}
\usepackage{dcolumn}
\usepackage{bm}
\usepackage{amsmath}
\usepackage{color}

\allowdisplaybreaks[2]

\begin{document}

\title{Control of beam propagation in optically written waveguides beyond the paraxial approximation}

\author{Lida Zhang}
\affiliation{Max-Planck-Institut f\"{u}r Kernphysik, Saupfercheckweg 1, D-69117 Heidelberg, Germany}

\author{Tarak N. Dey}
\affiliation{Department of Physics, Indian Institute of Technology Guwahati, Guwahati 781 039, Assam, India}
\affiliation{Max-Planck-Institut f\"{u}r Kernphysik, Saupfercheckweg 1, D-69117 Heidelberg, Germany}

\author{J\"{o}rg Evers}
\affiliation{Max-Planck-Institut f\"{u}r Kernphysik, Saupfercheckweg 1, D-69117 Heidelberg, Germany}

\date{\today}

\begin{abstract}
Beam propagation beyond the paraxial approximation is studied in an optically written waveguide structure.
The waveguide structure that leads to diffractionless light propagation, is imprinted on a medium consisting of a five-level atomic vapor driven by an incoherent pump and two coherent spatially dependent control and plane-wave fields. We first study propagation in a single optically written waveguide, and find that the paraxial approximation does not provide an accurate description of the probe propagation. We then employ coherent control fields such that two parallel and one tilted Gaussian beams produce a branched waveguide structure. The tilted beam allows selective steering of the probe beam into different branches of the waveguide structure. The transmission of the probe beam for a particular branch can be improved by changing the width of the titled Gaussian control beam as well as the intensity of the spatially dependent incoherent pump field.

\end{abstract}

\maketitle

\section{Introduction}

Diffractionless optical beam steering is of great interest in optical imaging, laser machining and optical communication. A gradient of the refractive index provides the simplest way to steer the optical beam.  Such a refractive index gradient can be achieved by various physical mechanisms such as thermal gradients \cite{Jackson}, acousto-optic interactions \cite{Dixon} and electro-optic effects \cite{Spencer}. Similarly, light-matter interactions can be used to control the optical beam propagation through the atomic medium. More precisely, electromagnetically induced transparency (EIT) \cite{Harris1} can provide beam-steering since refractive index changes significantly near the center of the transparency window \cite{Harris2, Sun}. 

Crucial obstacles in  the above schemes are absorption and diffraction. The effect of diffraction is inevitable as an optical beam propagates through a medium. An optical beam can be considered as a superposition of plane waves with different wave vectors. Diffraction arises, as each plane wave acquires a different phase during its propagation. In free space,  a light beam will typically be distorted severely already after propagation over few Rayleigh lengths~\cite{rayleigh}. For all-optical processing, this remains as a major obstacle in practical applications. 
In order to suppress or even eliminate diffraction, researchers have developed many proposals based on different physical mechanisms, such as EIT~\cite{moseley1,moseley2,truscott,kapoor,yavuz,gorshkov,li,howell}, coherent population trapping (CPT)~\cite{kapale,kiffner}, coherent Raman processes~\cite{dey1,vudyasetu}, or saturated absorption~\cite{dey2}. Most of the schemes employ suitable spatially-dependent structures of the control field to prevent the optical beam from diffracting. Recently, it was found that alternatively Dicke narrowing induced by atomic thermal motion and velocity-changing collisions can be useful to eliminate the diffraction of a probe beam that carries an arbitrary image~\cite{firstberg3,firstberg4}. 

A particular class of beam steering devices are so-called Y-branch waveguides, in which a single waveguide splits into two output ports~\cite{branch1,branch2,jia,sugimoto}.  Y-branch waveguides can be used to divide a single beam into two separate branches with a certain intensity ratio. A desirable feature of such devices is a dynamical control of the light beam intensity at the different branches.

Motivated by this, here we propose a scheme for all-optical beam steering in optically written waveguides. We facilitate a five-level medium driven by spatially dependent control fields, which was recently proposed as a method to achieve large refractive index modulations with low absorption~\cite{obrien}.
We start by analyzing the light propagation through a single optically written waveguide. Comparing results obtained using a split-operator method assuming paraxial approximation to those from a finite-difference-time domain approach beyond paraxial approximation, we find that already in this simple case the paraxial approximation usually assumed in related studies does not provide an accurate description of the beam dynamics. 
Next, we consider an optically written branched waveguide structure consisting of two parallel beams crossed by a third tilted beam. This structure essentially consists of two coupled Y-branched waveguides. We find that the tilted beam can be used to switch the light propagation between either of the two output ports formed by the two parallel light beams. Our numerical results show that the coupling efficiency of the probe beam into the different branches can be controlled by changing the width and the angle of the tilted Gaussian control field, and the transmission of the output probe beam can be improved by increasing the amplitude of an additionally applied incoherent pump field.

The article is organized as follows. In Sec.~\ref{sec2}, we introduce the theoretical model, analytically derive the linear response of the medium to the probe field from the master equation, and discuss the propagation equations and numerical methods to calculate within and beyond paraxial approximation. Sec.~\ref{sec3} describes the waveguide structure written by a Gaussian control field inside the atomic medium, and discusses the possibility of controlling the optically written waveguide structure by an incoherent pump field. In Sec.~\ref{sec4A}, we compare numerical results for the light propagation through a single waveguide within or beyond paraxial approximation. Sec.~\ref{sec4B} presents our main results on the controlled light propagation in a branched waveguide structure. Sec.~\ref{sec5} summarizes the results.

\begin{figure}[tp]
\centering\includegraphics[width=8cm]{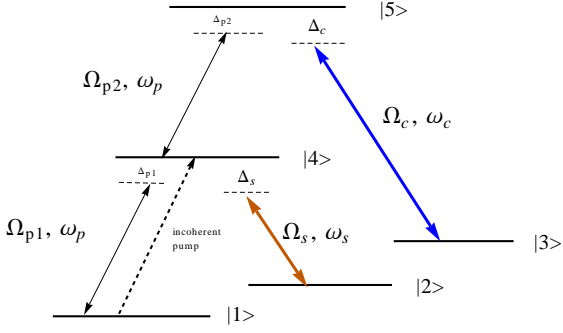}
\caption{The five-level type scheme considered in the analysis. The probe field couples to transitions $|1\rangle \leftrightarrow |4\rangle$ and $|4\rangle \leftrightarrow |5\rangle$ and The two coherent control laser fields $\Omega_{s},\Omega_{c}$ are far-detuned from respective transition frequencies in order to split states $|4\rangle$ and $|5\rangle$ into suitable dressed states. An incoherent pump control field is applied to $|1\rangle \leftrightarrow |4\rangle$ to achieve a population inversion. The control fields modify the probe field coupling to achieve high refractive index contrast with low absorption.}
\label{fig1}
\end{figure}

\section{\label{sec2}Theoretical Model}

\subsection{Equations of motion}
We consider the five-level atomic system shown in Fig.~\ref{fig1}. This model has been proposed previously in a different context for high index of refraction contrast with low absorption~\cite{obrien}. The two transitions $|2\rangle\leftrightarrow|4\rangle$ and $|3\rangle\leftrightarrow|5\rangle$ are driven by two far-detuned coherent laser fields with Rabi frequencies $\Omega_{s}$ and $\Omega_{c}$ respectively, and a weak probe is coupled to both transitions $|1\rangle\leftrightarrow|4\rangle$ and $|4\rangle\leftrightarrow|5\rangle$ with Rabi frequencies $\Omega_{p1}$ and $\Omega_{p2}$ respectively. An incoherent pump field is applied to transition $|1\rangle\rightarrow|4\rangle$ to provide a population inversion. 

The electric fields of the three beams are defined as
\begin{equation}
   \vec{\xi}_{j}(\vec{r},t)=\frac{1}{2}E_{j}(\vec{r},t)\:\vec{e}_{j}\:e^{-i\omega_{j}t+i\vec{k}\cdot\vec{r}}+\textrm{c.c.}\,,
\label{eq1}
\end{equation}
where $E_{j}(\vec{r},t)$ are the slowly varying envelopes and $\vec{e}_{j}$ the unit polarization vectors of the electric fields. The index $j\in\{p,s,c\}$ labels the three fields, respectively.
In dipole and rotating wave approximation, the Hamiltonian of the system in the interaction picture can be written as
\begin{align}
H_{I}/\hbar&=-(\Delta_{p1}-\Delta_{s})|2\rangle\langle2|-(\Delta_{p1}+\Delta_{p2}-\Delta_{s})|3\rangle\langle3| \nonumber \\
&\quad-\Delta_{p1}|4\rangle\langle4|-(\Delta_{p1}+\Delta_{p2})|5\rangle\langle5|-\left (\Omega_{p1}|4\rangle\langle1|\right.\nonumber \\
&\quad\left.+\Omega_{p2}|5\rangle\langle4|+\Omega_{s}|4\rangle\langle2|+\Omega_{c}|5\rangle\langle3|+ \textrm{H.c.}\right )\,,
\label{eq2}
\end{align}
where $\Delta_{p1}=\omega_{p}-\omega_{41}$, $\Delta_{p2}=\omega_{p}-\omega_{54}$, $\Delta_{s}=\omega_{s}-\omega_{42}$, and $\Delta_{c}=\omega_{c}-\omega_{53}$ are the detunings of the laser fields, and the Rabi frequencies of the fields are defined as   $\Omega_{p1}=E_{p}\vec{\mu}_{41}\cdot\vec{e}_{p}/2\hbar$, $\Omega_{p2}=E_{p}\vec{\mu}_{54}\cdot\vec{e}_{p}/2\hbar$, $\Omega_{s}=E_{s}\vec{\mu}_{42}\cdot\vec{e}_{s}/2\hbar$, and $\Omega_{c}=E_{c}\vec{\mu}_{53}\cdot\vec{e}_{c}/2\hbar$. Here,  $\vec{\mu}_{ij}$ are the dipole moments of the respective transitions, and we have simplified the notation $E_{j}(\vec{r},t)$ to $E_{j}$ $(j\in\{p,s,c\})$. The master equation of motion follows as
\begin{equation}
 \dot{\rho}=\frac{i}{\hbar}[H_{I},\rho]-\mathcal{L}\rho
\label{eq3}
\end{equation}
and $\mathcal{L}\rho$ represents the incoherent contributions given by
\begin{subequations}
\label{eq4}
\begin{align}
\mathcal{L}\rho &= \mathcal{L}_{41}^{\gamma}\rho+\mathcal{L}_{42}^{\gamma}\rho+\mathcal{L}_{53}^{\gamma}\rho+\mathcal{L}_{54}^{\gamma}\rho
+\mathcal{L}^{d}\rho + \mathcal{L}^{p}\rho
\,,
\\
\mathcal{L}_{jk}^{\gamma}\rho &= \frac{\Gamma_{jk}}{2}\left ( |j\rangle\langle j|\,\rho + \rho \,|j\rangle\langle j| - 2 |k\rangle\langle j|\,\rho\,|j\rangle \langle k|\right)\,,
\\
\mathcal{L}^{d}\rho &= \sum_{j\neq k} \gamma_{jk}^d\:|j\rangle\langle k|\,,
\\
\mathcal{L}^{p}\rho &= \frac{p}{2}\left ( |1\rangle\langle 1|\,\rho + \rho \,|1\rangle\langle 1| - 2 |2\rangle\langle 1|\,\rho\,|1\rangle \langle 2|\right)\,,
\end{align}
\end{subequations}
where $\mathcal{L}_{jk}^{\gamma}\rho$ describes spontaneous emission from $|j\rangle$ to $|k\rangle$ with rate $\Gamma_{jk}$. The second term $\mathcal{L}^{d}\rho$ models additional pure dephasing for $\rho_{jk}$ with rate $\gamma^{d}_{jk}$ such that the total damping rate of this coherence is $\gamma_{jk}=\gamma^{d}_{jk}+(\Gamma_{j}+\Gamma_{k})/2$, with $\Gamma_{j}=\sum_{k}\Gamma_{jk}$ the total decay rate out of state $|j\rangle$. The third contribution $\mathcal{L}^{p}\rho$ describes the incoherent pumping from $|1\rangle$ to $|4\rangle$ with rate $p$.

The equations of motion for the density matrix elements can be easily be derived to give
\begin{subequations}
\begin{align}
\dot{\rho}_{11}&=-p \rho _{11}-i \left(\Omega_{p1} \rho _{14}-\Omega_{p1} \rho _{41}\right)+\Gamma _{41} \rho _{44}\,,\\
\noalign{\medskip}
\dot{\rho}_{22}&=-i \Omega_{s}(\rho _{24}-\rho _{42})+\Gamma _{42} \rho _{44}\,,\\
\noalign{\medskip}
\dot{\rho}_{44}&=p \rho _{11}-\left(\Gamma _{41}+\Gamma _{42}\right) \rho _{44}-i \left(-\Omega_{p1} \rho _{14}-\Omega_{s} \rho _{24} \right.\nonumber \\
&\quad\left.+\Omega_{p1} \rho _{41}+\Omega_{s} \rho _{42}+\Omega_{p2} \rho _{45}-\Omega_{p2} \rho _{54}\right)+\Gamma _{54} \rho _{55}\,,\\
\dot{\rho}_{55}&=-i \left(-\Omega_{c} \rho _{35}-\Omega_{p2} \rho _{45}+\Omega_{c} \rho _{53}+\Omega_{p2} \rho _{54}\right) \nonumber \\
&\quad-\left(\Gamma _{53}+\Gamma _{54}\right) \rho _{55}\\
\noalign{\medskip}
\dot{\rho}_{42}&=i(\Delta_{s}+i \gamma _{42})\rho _{42}+i\Omega_{s}(\rho _{22}-\rho_{44})+i\Omega_{p1} \rho _{12}\nonumber \\
&\quad+i\Omega_{p2}\rho _{52}\,,\\
\noalign{\medskip}
\dot{\rho}_{54}&=-\gamma _{54} \rho _{54}-i \left(-\Omega_{c} \rho _{34}-\Omega_{p2} \rho _{44}+\Omega_{p1} \rho _{51}+\Omega_{s} \rho _{52}\right.\nonumber\\
&\quad\left.-\Delta_{p2} \rho _{54}+\Omega_{p2} \rho _{55}\right)\,,\\
\dot{\rho}_{41}&=-\left(p/2+\gamma _{41}\right) \rho _{41}-i \left(-\Omega_{p1} \rho _{11}-\Omega_{s} \rho _{21}-\Delta_{p1} \rho _{41}\right.\nonumber\\
&\quad\left.+\Omega_{p1} \rho _{44}-\Omega_{p2} \rho _{51}\right)\,,\\
\noalign{\medskip}
\dot{\rho}_{34}&=-\gamma _{43} \rho _{34}-i \left(\Omega_{p1} \rho _{31}+\Omega_{s} \rho _{32}+(\Delta_{c}-\Delta_{p2}) \rho _{34}\right.\nonumber\\
&\quad\left.+\Omega_{p2} \rho _{35}-\Omega_{c} \rho _{54}\right)\,, \\
\noalign{\medskip}
\dot{\rho}_{52}&=-\gamma _{52} \rho _{52}-i \left\{-\Omega_{c} \rho _{32}-\Omega_{p2} \rho _{42}-(\Delta_{s}+\Delta_{p2}) \rho _{52}\right.\nonumber\\
&\quad\left.+\Omega_{s} \rho _{54}\right\}\,,\\
\noalign{\medskip}
\dot{\rho}_{21}&=-\left(p/2+\gamma _{21}\right) \rho _{21}-i \left\{(-\Delta_{p1}+\Delta_{s}) \rho _{21}+\Omega_{p1} \rho _{24}\right.\nonumber\\
&\quad\left.-\Omega_{s} \rho _{41}\right\}\,,\\
\noalign{\medskip}
\dot{\rho}_{32}&=-\gamma _{32} \rho _{32}-i \left\{(\Delta_{c}-\Delta_{s}-\Delta_{p2}) \rho _{32}+\Omega_{s} \rho _{34}\right.\nonumber\\
&\quad\left.-\Omega_{c} \rho _{52}\right\}\,.
\end{align}%
\label{eq5}
\end{subequations}
The remaining equations follow from the constraints $\sum_{i}\rho_{ii}=1$ and $\rho_{ij}=\rho^\ast_{ji}$.

\subsection{Steady state solution}

We assume the probe field to be weak enough to be treated as a perturbation to the system in linear order. The related zeroth [superscript $(0)$] and first-order [superscript $(1)$] contributions for $\rho_{ij}$ are obtained as
\begin{subequations}
\begin{align}
\rho^{(0)}_{11}&=\frac{2 \gamma _{42} \Gamma _{41} \Omega _s^2}{p \gamma _{42}^2 \Gamma _{42}+p \Gamma _{42} \Delta _s^2+2 \gamma _{42} \left(2 p+\Gamma _{41}\right) \Omega _s^2}
\,,\\
\rho^{(0)}_{22}&=\frac{p \left(\gamma _{42}^2 \Gamma _{42}+\Gamma _{42} \Delta _s^2+2 \gamma _{42} \Omega _s^2\right)}{p \gamma _{42}^2 \Gamma _{42}+p \Gamma _{42} \Delta _s^2+2 \gamma _{42} \left(2 p+\Gamma _{41}\right) \Omega _s^2} \,, \\
\noalign{\medskip}
\rho^{(0)}_{44}&=\frac{2 p \gamma _{42} \Omega _s^2}{p \gamma _{42}^2 \Gamma _{42}+p \Gamma _{42} \Delta _s^2+2 \gamma _{42} \left(2 p+\Gamma _{41}\right) \Omega _s^2}\,,\\
\noalign{\medskip}
\rho^{(0)}_{55}&=0\,,\\
\rho^{(1)}_{41}&=\frac{\left(\rho^{(0)}_{44}-\rho^{(0)}_{11}\right) \Omega _{p1}}{\Delta _{p1}+i(\gamma_{41}+\frac{p}{2})+\frac{\Omega _s^2}{\Delta _{s}-\Delta _{p1}-i(\gamma_{21}+p/2)}}\,,\\
\noalign{\medskip}
\rho ^{(1)}_{54}&= \frac{A\left(\rho^{(0)}_{44}-\rho^{(0)}_{55}\right)\Omega _{p2}}{B}
\,.\end{align}
\label{eq6}
\end{subequations}
The detailed expressions for $A, B$ are provided in Appendix~\ref{app-ab}. 

\subsection{Interpretation of the level scheme}
To elucidate the role of the different couplings, we reduce the level structure to a three-level ladder system, by switching off  the two far-detuned laser fields. Then, the linear susceptibility for the probe simplifies to
\begin{align}
 \chi(\omega_{p})=\frac{3N\lambda^{3}_{p}}{8\pi^{2}}\left(\frac{\Gamma_{41}(\rho_{44}-\rho_{11})}{\Delta _{p1}+i(\gamma_{41}+\frac{p}{2})}+ \frac{\Gamma_{54}(\rho_{55}-\rho_{44})}{\Delta _{p2}+i\gamma_{54}}\right)\,,
\label{eq7}
\end{align}
where $N$ is the atomic density and $\lambda_{p}$ is the wave length of the probe field. The terms in the right hand side of Eq.~(\ref{eq7}) describe the two individual contributions coming from the two two-level subsystems in the three-level ladder structure.  To achieve the desired refractive index modulation without absorption, three conditions should be satisfied in Eq.~(\ref{eq7})~\cite{obrien}. First, the population inversions on the two transitions should be matched. Second, the decoherence rates should be comparable. Third, the two-photon resonance should be fulfilled for the two transitions. 
These conditions can be expressed as $\Gamma_{41}(\rho_{44}-\rho_{11})=\Gamma_{54}(\rho_{55}-\rho_{44})$, $\gamma_{54}=\gamma_{41}+p/2$ and $\Delta_{p1}+\Delta_{p2}=0$.  A maximum refractive index modulation can then be realized if $\gamma_{41}=\pm\Delta_{p1}-p/2$. In this ideal case, the refractive index is increased by a factor of 2, with absorption canceled at the same time. 
However, it is hard to find a real atomic system in which these conditions are satisfied simultaneously. To relax the stringent conditions, it has been suggested in~\cite{obrien} to introduce the two far-detuned laser fields coupling $|4\rangle$ and $|5\rangle$ to auxiliary states to induce Stark shifts. The Stark shifts of $|4\rangle$ and $|5\rangle$ together with modifications to the decoherence rates of the Stark sublevels can be controlled by tuning the intensities and detunings of the two laser fields. This way, the two transitions for the probe can be modified to one's advantage. This leads to the five-level scheme shown in Fig.~\ref{fig1}, in which the two upper states are coupled by two external far-detuned laser fields $\Omega_{s},\Omega_{c}$. By appropriately choosing the parameters of these laser fields, one can achieve strong refractive index modulation with minimized absorption or even gain even in realistic level schemes.

\subsection{Propagation dynamics with and without paraxial approximation}
The propagation dynamics of the probe field is governed by  Maxwell's equations. Without applying the paraxial approximation, the wave equation for the probe field propagating along $z$ direction can be derived as~\cite{goodman}
\begin{equation}
 \frac{\partial E_{p}}{\partial z}=\frac{ic}{2\omega_{p}}(\nabla_{\bot}^2+\frac{\partial^{2}}{\partial z^{2}})E_{p}+\frac{ik_{p}}{2}\chi(\omega_{p})E_{p}\,,
\label{eq12}
\end{equation}
where $k_{p}=\omega_{p}/c$ with $c$ the speed of light in vacuum and $\nabla_{\bot}^{2}=\partial^{2}/{\partial x}^{2}+\partial^{2}/{\partial y}^{2}$. In Eq.~(\ref{eq12}), the diffraction of the probe, which is induced by the term $ic\nabla_{\bot}^2 E_{p}/2\omega_{p}$, may be reduced or even eliminated by the other term $ik_{p}\chi(\omega_{p})E_{p}/2$ if the real part of the susceptibility has a waveguide-like structure. 

To evaluate Eq.~(\ref{eq12}) without the paraxial approximation, we make use of the finite-difference time domain (FDTD) technique~\cite{taflove} provided by the software package MEEP~\cite{johnson}. In order to compare the results to those obtained in paraxial approximation, we drop the term $(ic/2\omega_{p})\partial^{2}E_{p}/\partial z^{2}$ which introduces effects beyond the paraxial approximation, and solve the resulting reduced equation using the Strang split operator method~\cite{strang}.  In both simulation techniques, all fields are chosen as continuous wave fields.

\section{\label{sec3}Realizing spatial waveguide-like structures}
\subsection{Basic analysis}
\begin{figure}[t]
\centering\includegraphics[width=8cm]{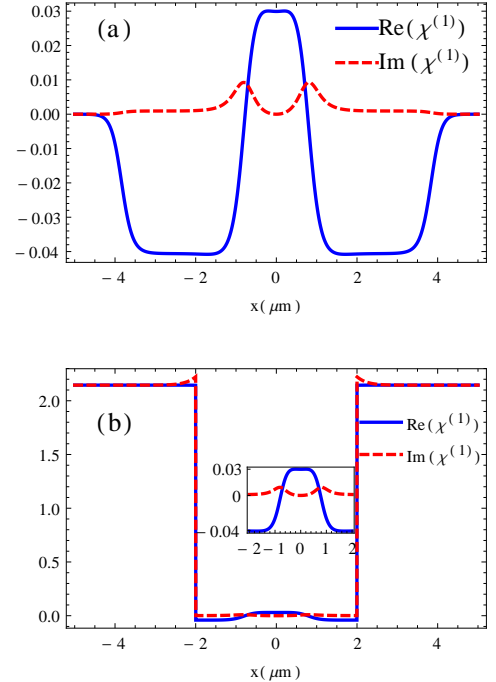}
\caption{(Color online) Real (blue solid line) and imaginary (red dashed line) part of the linear susceptibility obtained from Eq.~(\ref{eq8}) as a function of the position in the presence of a control laser field $\Omega_{s}$ with Gaussian intensity profile. Panel (a) shows results for a constant incoherent pump rate, whereas (b) shows results for spatially dependent pumping as explained in the main text. The inset in (b) depicts the detailed structure of the linear susceptibility in the central area which is similar to that in (a). Note that the profile of the real part in the central region forms a waveguide structure. Parameters are chosen as: $N=1.4\times10^{27}$~m$^{-3}$, $\lambda_{p}=813.2$~nm, $\gamma^{d}_{54}=0.8$~GHz, $\gamma^{d}_{41}=\gamma^{d}_{42}=0.3$~GHz, $\gamma^{d}_{43}=0.2$~GHz, $\gamma^{d}_{52}=\gamma^{d}_{21}=\gamma^{d}_{32}=0$, $\Gamma_{41}=\Gamma_{42}=45$~Hz, $\Gamma_{54}=15$~Hz, $\Delta_{p1}=-\gamma_{41}$, $\Delta_{p2}=19.7$~GHz, $\Delta_{s}=10.0$~GHz, $\Delta_{c}
=18.0$~GHz, $\Omega_{s0}=\sqrt{2\gamma_{41}\Delta_{s}}$, $\Omega_{c}=5.2$~GHz, $w_{s}=1.0$~$\mu$m, and $p=45.9943$~Hz.}
\label{fig2}
\end{figure}

From Eqs.~(\ref{eq6}), the linear susceptibility of the probe field can be written as
\begin{equation}
 \chi^{(1)}=\frac{3N\lambda^{3}_{p}}{8\pi^{2}}\left (\frac{\Gamma_{41}\rho^{(1)}_{41}}{\Omega_{p1}}+\frac{\Gamma_{54}\rho^{(1)}_{54}}{\Omega_{p2}}\right )\,.
\label{eq8}
\end{equation}
We now assume a control field with spatial intensity dependence modeled as a Gaussian 
\begin{align}
\Omega_{s}=\Omega_{s0}\:e^{-\frac{x^2}{2w_{s}^2}}\,,
\label{eq9}
\end{align}
where $w_{s}$ is the width of the intensity profile. This spatial dependence of the control field creates a modulation of the medium  susceptibility in space, which can be used as a waveguide. 
In order to achieve maximum refractive index modulation, the peak Rabi frequency of the laser field $\Omega_{s}$ should meet the condition $\Omega_{s0}=\sqrt{2\gamma_{41}\Delta_{s}}$. Assuming these conditions, an example for the real and imaginary part of the susceptibility, which correspond to the medium dispersion and absorption, respectively, is shown in Fig.~\ref{fig2}(a).
Clearly, the real part of the susceptibility resembles a waveguide structure, whereas absorption is low. 

The parameters in Fig.~\ref{fig2} are chosen to fulfill  $\Delta_{p1},\Delta_{s}, \gamma_{54},\gamma_{41}\gg \Gamma_{54},\Gamma_{41},p$. In this limiting case,  Eq.~(\ref{eq6}e) can be simplified to
\begin{equation}
\rho^{(1)}_{41}=\frac{\left(\rho^{(0)}_{44}-\rho^{(0)}_{11}\right) \Omega _{p1}}{-\gamma_{41}+i\gamma_{41}+\frac{\Omega _s^2}{\Delta _{s}+\gamma_{41}}}\,.
\label{eq10}
\end{equation}
In the central area of $\Omega_{s}$ defined by Eq.~(\ref{eq9}), the part $\Omega_{s}^2/(\Delta_{s}+\gamma_{41})-\gamma_{41}\simeq\gamma_{41}$  in the dominator of Eq.~(\ref{eq10}), and the refractive index is maximized as desired together with gain. At the same time, the transition $|5\rangle\leftrightarrow|4\rangle$ will give maximum refractive index together with absorption. Those two transitions together result in a maximum refractive index for the probe with almost canceled absorption in the central area of $\Omega_{s}$ as shown in Fig.~\ref{fig2}(a). 
In contrast,  in the two wings of $\Omega_{s}$, the refractive index is minimized for transition $|4\rangle\leftrightarrow|1\rangle$ with gain since $\Omega_{s}^2/(\Delta_{s}+\gamma_{41})-\gamma_{41}\simeq-\gamma_{41}$. Simultaneously, a minimal refractive index with absorption is obtained for transition $|5\rangle\leftrightarrow|4\rangle$. In total, a minimal refractive index with little absorption is obtained for the probe. 

In order to check the validity of the perturbation approximation in the far-detuned case, we also obtained the probe field susceptibility to all orders by numerically solving Eq.~(\ref{eq5}) in steady state. For a weak probe field $\Omega_{p1}\sim\Gamma_{41}$, it is virtually indistinguishable from the linear susceptibility shown in Fig.~\ref{fig2}(a).

\subsection{Incoherent pumping}
In this section, we discuss the possibility to control the optically written waveguide structure by means of an incoherent pump field. As shown in Fig.~\ref{fig2}(a), the absorption becomes weaker in the two wings than in the central area. In practical implementations, this may lead to stray fields outside the waveguide region caused by the part of the probe field leaking out of the waveguide-like structure in the central part into the weakly absorbing wings. In order to eliminate this leaking light, we propose a spatially dependent incoherent pump field given as follows
\begin{equation}
p(\Omega_{s}) = \left\{
\begin{array}{l l}
p & \quad \text{if} \quad (\frac{\Omega_{s}}{\Omega_{s0}})^2 \geqslant e^{-4}\\
\\
0 & \quad \text{if} \quad (\frac{\Omega_{s}}{\Omega_{s0}})^2 < e^{-4}
\end{array} \right.\,.
\label{eq11}
\end{equation}
The resulting linear susceptibility is shown in Fig.~\ref{fig2}(b), for otherwise same parameters as in (a). It can be seen that the   central region featuring the waveguide-like structure remains the same as in Fig.~\ref{fig2}(a) as expected. In contrast, in both side wings, where $(\Omega_{s}/\Omega_{s0})^2<e^{-4}$, the incoherent pump vanishes, such that the atoms remain in the ground state $|1\rangle$ according to Eq.~(\ref{eq6})(a). Thus, the medium  acts as an absorber, and stray-fields outside the waveguide region are reduced. In the following, the spatially-dependent incoherent pump is used in the numerical simulations.

\section{\label{sec4}Results}

\subsection{\label{sec4A}Validity of the paraxial approximation}

\begin{figure}[t]
\centering\includegraphics[width=9cm]{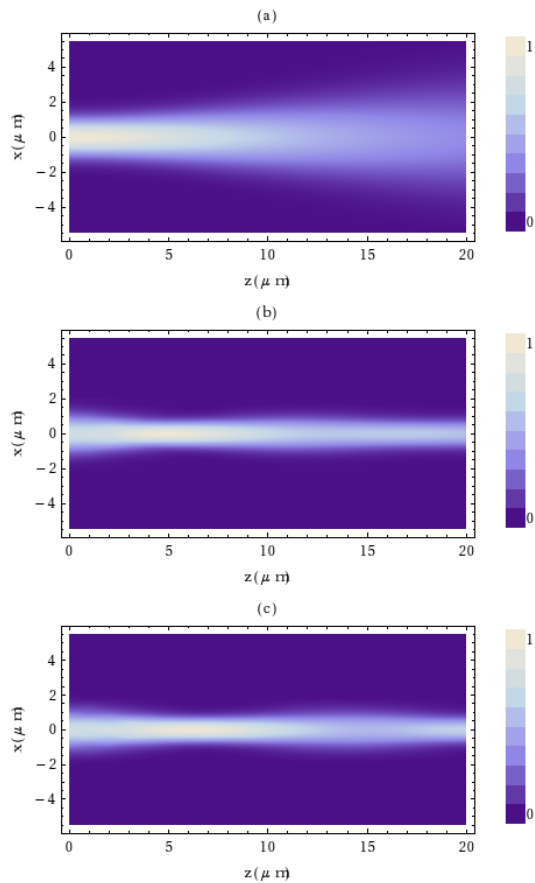}
\caption{(Color online) Normalized intensity of the probe field as a function of propagation distance $z$. (a) shows the free space case, (b) the propagation in a medium prepared with a control field and evaluated beyond the paraxial approximation, and (c) the corresponding results evaluated in paraxial approximation. The maximum propagation distance is $z_{0}=20\mu$m. The initial width of the probe is $w_{p}=1.0\mu$m at $z=0$. Other parameters are as in Fig.~\ref{fig2}.}    
\label{fig3}
\end{figure}

\begin{figure}[t]
\centering\includegraphics[width=8cm]{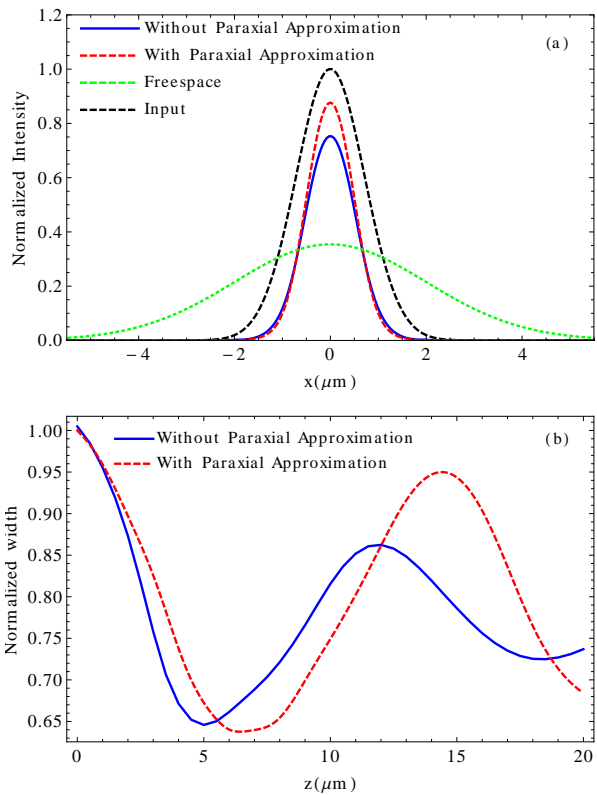}
\caption{(Color online) (a) Intensity profile of the propagating probe field at the medium entrance (dashed black line) and output after a propagating distance $z_{0}=20~\mu$m. The dotted green line shows propagation in free space, the solid red line propagation in the medium with control field evaluated in paraxial approximation, and the solid blue line corresponding results beyond paraxial approximation. (b) Transversal width of the probe field as a function of propagation distance $z$ evaluated within (dashed red) and beyond (solid blue) paraxial approximation. Parameters are chosen as in Fig.~\ref{fig3}.}
\label{fig4}
\end{figure}

As a first step, we study two-dimensional (2D) light propagation of a Gaussian  probe field along $z$. The probe field width is $w_{p}=1.0\mu$m at $z=0$. The medium is prepared by a Gaussian control laser field  $\Omega_{s}$ with width $w_{s}=1\mu$m at $z=0$. The respective beam profiles are as in Eq.~(\ref{eq9}). Note that we do not take into account the spreading of the field $\Omega_{s}$. Fig.~\ref{fig3} shows results for the normalized intensity of the probe at different positions for a propagation distance of $z_{0}=20\mu$m (about 2.59 Rayleigh lengths $z_{R}=2\pi w_{p}^{2}/\lambda_{p}\simeq7.73\mu$m). The three panels compare different cases. Panel (a) shows the free-space case without control field. Clearly, the diffraction spreading of the probe beam can be seen. In (b), results are shown for the FDTD simulation beyond the paraxial approximation. It can be seen that the diffraction is suppressed by the waveguide structure introduced by the control beam. Finally, in (c), results in paraxial approximation from the Strang propagation technique are shown. While again a waveguide-like propagation without diffraction, surprisingly, the results differ considerably from those obtained in (b) beyond paraxial approximation. In particular, the paraxial approximation leads to higher maximum intensity, and the ``breathing''-like width modulation throughout the propagation is slower than in (b).\\
\indent Fig.~\ref{fig4} shows corresponding sections through Fig.~\ref{fig3}, depicting the intensity profile for the different cases at propagation distance $z_0$. In (a), it can be seen that in free space, the maximum intensity is considerably attenuated together with a substantial increase in the width to $w_{p}(z_{0})\simeq2.85\mu$m. With control field beyond the paraxial approximation, the width is slightly narrowed compared to the input width to $w_{p}(z_{0})\simeq0.75\mu$m, together with a moderate reduction of the maximum intensity. In paraxial approximation, the width is under- and the maximum intensity is over-estimated.\\
\indent An intuitive way to understand the observed diffractionless light propagation is to model the probe as an ensemble of rays propagating in a medium rendered by the applied control field into a waveguide structure. As the probe is propagating in the medium, total reflection at the medium boundaries occurs, as the refractive index is larger in the central area than in the two wings as shown by the solid blue line in Fig.~\ref{fig2}(a). Thus, the main part of the probe intensity is confined in the central area. To substantiate this interpretation, we calculated the width of the probe as a function of propagation distance $z$, and show the result in Fig.~\ref{fig4}(b). It can be seen that the width of the probe oscillates periodically throughout the propagation. These oscillations in the width arise from the total refraction. Those rays in the probe which do not meet the condition for total internal reflection leave the central area and are absorbed in the wings. This leads to the reduction in both the 
energy density and the width of the probe as the propagation distance increases. Note that Fig.~\ref{fig4}(b) again shows the difference in the spatial structure of the probe beam with and without the paraxial approximation.\\
\indent We thus conclude, that even when confined to a waveguide like structure and when propagating only few Rayleigh lengths, the paraxial approximation does not provide an accurate description of the probe field propagation.\\

\begin{figure}[tp]
\centering\includegraphics[width=10cm]{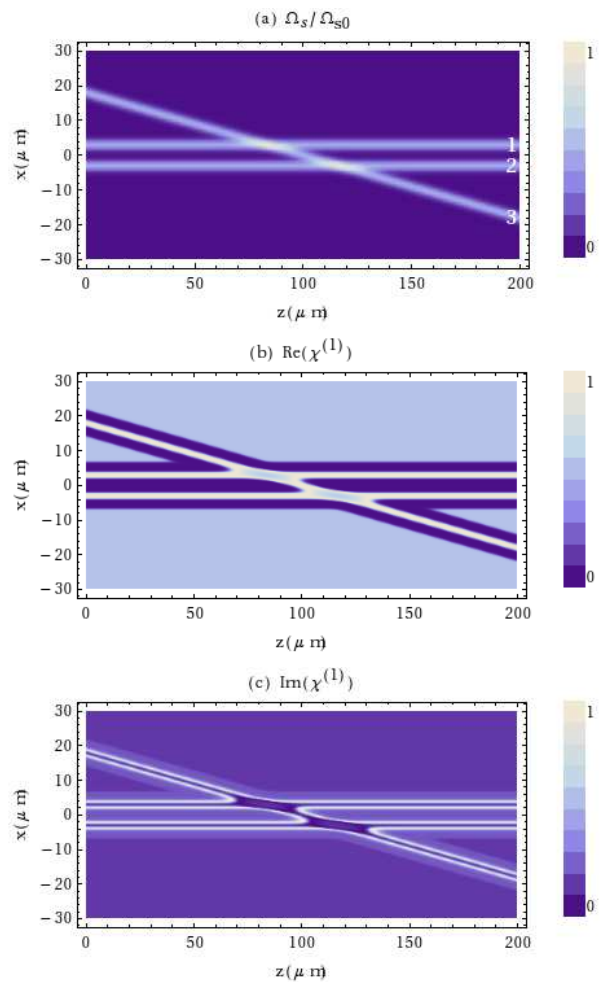}
\caption{(Color online) Sketch of the branched waveguide structure. (a) shows the spatial profile of $\Omega_{s}$ for $k=1$. Panels (b) and (c) show the corresponding position-dependent real and imaginary part of the linear susceptibility in the medium. The labels $i\in\{1, 2, 3\}$ in (a) indicate the three output ports corresponding to the $i$-th control beam $\Omega_{si}$. The size of the medium is chosen as $60$~$\mu$m~$\times$~$200$~$\mu$m. The other parameters are as in Fig.~\ref{fig2}.}
\label{fig5}
\end{figure}

\subsection{\label{sec4B}All-optical switching of light in a branched waveguide structure}
We now turn to our main results, on applying the waveguide-like structures obtained in the previous section for all-optical switching. For this, we consider light propagation in a more complicated branched waveguide structure formed by a field $\Omega_{s}$ consisting of three Gaussian laser beams with different geometries defined as
\begin{subequations}
\begin{align}
\Omega_{s}&=\Omega_{s1}+\Omega_{s2}+\Omega_{s3}\,,\\
\Omega_{s1}&=\Omega_{s0}e^{-\frac{(x-x_{0})^2}{2w_{s}^2}}\,, \\
\Omega_{s2}&=\Omega_{s0}e^{-\frac{(x+x_{0})^2}{2w_{s}^2}}\,,\\
\Omega_{s3}&=\Omega_{s0}e^{-\frac{[(x-x_{t})\cos\theta+z\sin\theta]^2}{2k^2w_{s}^2}}\,.
\end{align}
\label{eq13}
\end{subequations}
The spatial profile of the field $\Omega_{s}$ is plotted in Fig.~\ref{fig5}(a). It consists of two parallel beam structures, intersected by a tilted one. In this figure, the width $w_{s}$, the displacement $x_{0}$ of the parallel beams, as well as the displacement $x_{t}$ of the tilted beam  are chosen as $1.0$~$\mu$m, $3.0$~$\mu$m, and $18.0$~$\mu$m, respectively. $k$ is a factor by which the width of the tilted control beam differs from that of the parallel beams. The maximum propagation distance is $z_{0}=200~\mu$m, and the tilting angle for $\Omega_{s3}$ is $\tan\theta=2x_{t}/z_{0}$. Using Eqs.~(\ref{eq6}), (\ref{eq8}) and (\ref{eq13}), we can obtain the linear susceptibility for the probe, which has real and imaginary parts as shown in Fig.~\ref{fig5}(b) and \ref{fig5}(c), respectively. It can be seen that branched waveguide-like structures are generated in the optical response corresponding to the three beams of $\Omega_{s}$. Note that in regions in which two of the fields $\Omega_{s1}$, $\Omega_{s2}$ 
and $\Omega_{s3}$ overlap, the total magnitude of $\Omega_{s}$ is considerably  larger than $\Omega_{s0}$ of a single field, which means that the conditions for the desired maximum reflective index modulation are not satisfied. This deviation leads to lower spatial dispersion with weak gain as shown in the corresponding regions in Figs.~\ref{fig5}(b) and \ref{fig5}(c). As a consequence, refraction will take place when the probe field enters these regions. 

We found that the branched waveguide structure can be used to guide the probe light selectively into either of the three output ports on the right hand side. For this, we consider a Gaussian probe field which is launched into the waveguide formed by $\Omega_{s1}$ left of the intersection point with the tilted control beam. As a first step, we have calculated the light propagation without the tilted control field ($\Omega_{s3}=0$), such that the probe light remains in the upper of the two parallel waveguide structures. We denote this transmitted power after propagation of a single optically written waveguide as $P_{0}$. We will normalize part of the results of the following analysis to this reference value $P_0$, and it is important to remember that $P_0$ is smaller than the incident power due to attenuation in the waveguide.
ions.

We then switch on the tilted control beam, and calculate the field power at the three output ports after the propagation, which we denote as $P_{i}$. From these quantities, we calculate the relative transmitted powers $T_{i}=P_{i}/P_{0}$. Here, $i=1$ corresponds to the upper parallel waveguide, $i=2$ to the lower parallel waveguide, and $i=3$ to the tilted waveguide.

\begin{figure}[t]
\centering
\setlength{\dbltextfloatsep}{5mm}
\includegraphics[width=8cm]{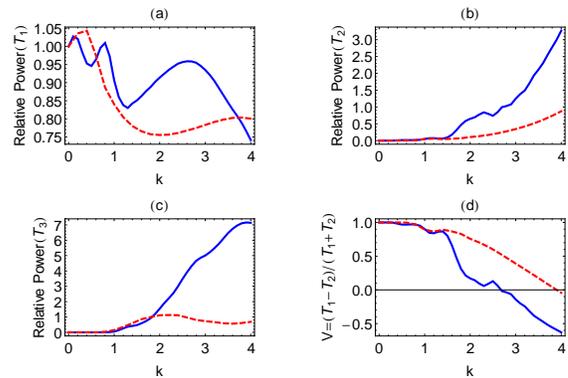}
\caption{(Color online) (a,b,c) Transmitted powers $T_{1}$, $T_{2}$, and $T_{3}$ at the three output ports indicated in Fig.~\ref{fig5}(a) after the propagation through the branched waveguide medium. (d) shows the visibility $V=(T_{1}-T_{2})/(T_{1}+T_{2})$ for the switching of light between the two output ports $1$ and $2$. All results are plotted against the relative width of the tilted control beam  $k$. The blue solid (red dashed) line shows results for tilting angle $\tan\theta=0.18$ ($\tan\theta=0.26$). The probe beam initially has a Gaussian shape with $w_{p}=1.0$~$\mu$m. Other parameters are as in Fig.~\ref{fig2}.}
\label{fig6}
\end{figure}

\begin{figure}[t]
\centering
\setlength{\dbltextfloatsep}{5mm}
\includegraphics[width=8cm]{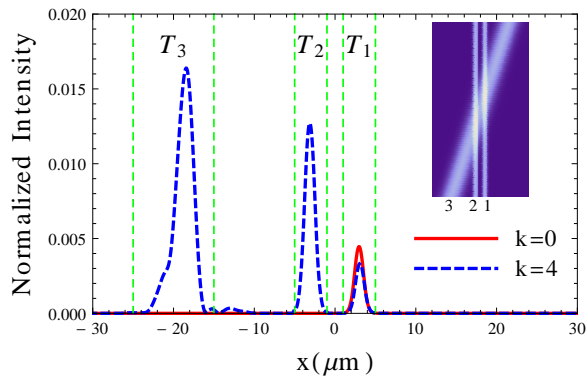}
\caption{(Color online) Spatial field configuration in the plane transverse to the propagation direction at the output of the medium at $z_{0}=200~\mu$m. The red solid line shows results without tilted control beam, whereas the blue dashed line shows results including a tilted control with relative width $k=4$. The vertical green dashed lines indicate the positions of the three output ports.  The tilting angle is chosen as $\tan\theta=0.18$. Other parameters are as in Fig.~\ref{fig6}.}
\label{fig7}
\end{figure}

Results are shown in Fig.~\ref{fig6}(a-c) for two different tilting angles $\tan \theta \in\{0.18, 0.26\}$, as a function of the tilted beam width $k$. Additionally, in (d), the visibility $V=(T_{1}-T_{2})/(T_{1}+T_{2})$ is shown, which can be seen as a figure of merit for the switching between the two parallel output ports. While for $k\to 0$, all light is emitted at output port 1 as expected, with increasing $k$, the light is re-routed towards port 2, with a visibility below $-0.6$. This indicates that switching of the output port by means of the tilted control field is possible. 
The corresponding output probe field structure after propagating through the branched waveguide structure without tilted field ($k=0$) and with $k=4$ is shown in Fig.~\ref{fig7}. It clearly shows that the probe pulse in output 2 exceeds that in port 1. 

Fig.~\ref{fig7} also shows that the output power can be larger than $P_{0}$. This is also shown in Figs.~\ref{fig6}(a-c). The origin of this increase is the slight gain in the field overlap regions. It should be noted, however, that overall the sum of the total transmitted probe intensity in all three output ports is lower than the input power. Values larger than $P_0$ only indicate that the transmitted power is larger than the power transmitted through a single optically written waveguide, as part of the absorption is compensated by the gain in the overlap region.

In the following, we provide a simple explanation for the observed light switching to different output ports. If the tilted beam is switched off, the probe field propagates without diffraction to output port 1 as discussed in the previous section.  When the tilted beam $\Omega_{s3}$ is turned on, and if suitable parameters are chosen for the tilting angle, the tilted beam width, and the atomic density $N$,  the probe field can be refracted into the tilted beam $\Omega_{s3}$. To understand this, on should note that the region in which two of the three control fields overlap has a real part of the susceptibility varying with the propagation direction $z$. Due to this variation, as well as due to the missing guiding in perpendicular direction in the overlap region, parts of the probe field can be redirected into the tilted beam. Starting from a tilted beam of vanishing thickness, initially the proportion of redirected light increases with increasing thickness. After the first intersection point, at the second intersection point, in turn light can be redirected out of the tilted beam into the lower parallel control beam.

Note that the transmitted power $T_{1}$ at output port 1 oscillates with the thickness of the tilted control beam determined by $k$. This is likely an interference effect. As discussed above, the refractive index varies along the propagation direction $z$ in the waveguide formed by $\Omega_{s1}$ in the intersection point. When the probe field enters this area, some part of it is refracted back and forth repeatedly along the $z$ direction. Depending on $k$, the interference of the different channels possible for the light changes between (partly) constructive or destructive, thus resulting in the oscillations in $T_{1}$.

\begin{figure}[t]
\centering
\setlength{\dbltextfloatsep}{5mm}
\includegraphics[width=8cm]{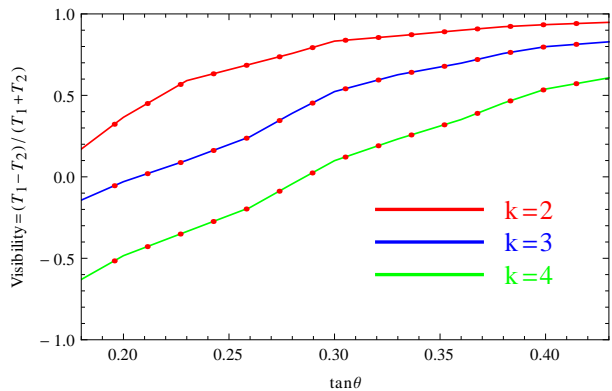}
\caption{(Color online) Visibility $V=(T_{1}-T_{2})/(T_{1}+T_{2})$ as a function of the tilting angle of the control field for three different relative widths of the tilted beam. The upper red line shows $k=2$, the middle blue line $k=3$, and the lower green line $k=4$. Other parameters are as in Fig.~\ref{fig6}.}
\label{fig8}
\end{figure}

\begin{figure}[t]
\centering
\setlength{\dbltextfloatsep}{5mm}
\includegraphics[width=8cm]{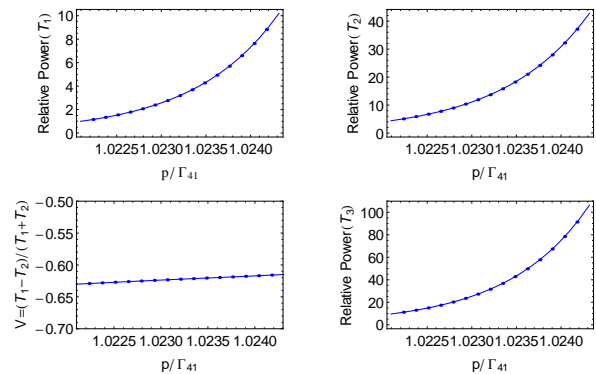}
\caption{(Color online) The effect of the incoherent pump on the output of the probe. The angle and width of the probe are chosen as $\tan\theta=0.18$ and $k=4$ respectively. Other parameters are as in Fig.~\ref{fig6}.}
\label{fig9}
\end{figure}

Qualitatively, one would expect that the probe field can more easily be redirected from $\Omega_{s1}$ into the tilted beam $\Omega_{s3}$ and subsequently into the lower parallel guide $\Omega_{s2}$ if the tilting angle $\theta$ is smaller, and if the atomic density $N$ is larger. We have verified this by plotting the transmitted power $T_{i}$ and visibility $V$ versus the width of $\Omega_{s3}$ also for a larger tilting angle $\tan\theta=0.26$ in Fig.~\ref{fig6}. It can be seen that the probe transferred into the two output ports 2 and 3 is generally smaller at $\tan\theta=0.26$ as compared to that at $\tan\theta=0.18$. We then further calculated the visibility V as a function of the tilting angle $\tan\theta$ for several different widths of the tilted control. results are shown in Fig.~\ref{fig8}. It can be seen that $V$ increases as the tilting angle $\theta$ increases for all widths of the tilted beam.

It should be noted that since the tilting angle in our case is relatively large ($\tan\theta \in \{0.18, 0.26\}$), the paraxial approximation cannot be applied. This is clearly indicated by the non-vanishing transmitted power $T_{3}$ which arises due to effects beyond the paraxial approximation described by the term $(ic/2\omega_{p})\partial^{2}E_{p}/\partial z^{2}$ in Eq.~(\ref{eq12}).

As discussed in Sec.~\ref{sec4A}, since we have chosen a relatively weak incoherent pump, the probe will be absorbed as it propagates in the medium. This is an obstacle in experimentally realizing the proposed scheme. However, the output of the probe can be improved by applying stronger incoherent pumping. Fig.~\ref{fig9} shows the transmitted power $T_{i}(i=1,2,3)$ and visibility $V$ as a function of the incoherent pump rate $p$. It can be seen that the transmitted powers $T_{i}$ can be sensitively controlled via the intermittent gain introduced via $p$ over a wide range of output powers. Note that due to the slightly longer propagation distance to output port 2 compared to the path to 1, the transmitted power  $T_{2}$ grows faster than  $T_{1}$ as the incoherent pump increases. For the same reason, $T_{3}$ grows even more rapidly. 
It is important to note that while the transmitted powers can be controlled, the visibility for the switching remains approximately  the same over the whole range of $p$. This suggests that the visibility is mainly determined by the geometry of the optically written structure, consistent with the waveguide interpretation. In contrast to the absorption and gain properties, this geometry is largely independent of the pumping $p$, with the exception of slight changes in the beam profiles with $p$.

\section{\label{sec5}Conclusion}

We have investigated light propagation in an optically written waveguide structure. The waveguide structure is prepared in a medium tailored with coherent control fields and an incoherent pump field. A spatially dependent Gaussian control field is applied to create a waveguide-like spatial refractive index variation within the medium, such that a probe beam can propagate within the waveguide structure essentially without diffraction. Our initial calculations show that already in a single optically written waveguide, accurate results for the beam propagation cannot be obtained within the paraxial approximation. To analyze the propagation dynamics, we therefore numerically integrate  the corresponding equations without paraxial approximation for the probe fields. 

As our main result, we have demonstrated that a controllable branched-waveguide structure can be optically formed inside the medium by applying a spatially dependent control field consisting of two parallel and one tilted Gaussian beams. The tilted beam can be used to selectively steer a probe beam between two different output ports. In the overlapping regions of the Gaussian control beams a rapidly varying refractive index is created that guides the probe beam to propagate in a particular output branch. Increasing the amplitude of the incoherent pump beam can be used to reduce the absorption of the medium.

\begin{acknowledgements}
We are grateful for funding by the German Science Foundation
(DFG, Sachbeihilfe EV 157/2-1).  T.N.D. gratefully acknowledges funding by the Science and Engineering Board (SR/S2/LOP-0033/2010).
\end{acknowledgements}

\appendix

\section{\label{app-ab}Appendix}
The explicit expressions for $A,B$ in Eqs.~(\ref{eq6}) are as follows:
\begin{widetext}
\begin{subequations}
\begin{align}
A&=-i \left(-\gamma _{32}-i \left(\Delta _c-\Delta _{p2}-\Delta _s\right)\right) \left(\left(-i \gamma _{32}+\Delta _c-\Delta _{p2}-\Delta _s\right) \left(i \gamma _{52}+\Delta _{p2}+\Delta _s\right)+\Omega _c^2-\Omega _s^2\right) \biggl(\gamma _{43}+i \left(\Delta _c-\Delta _{p2}\right)+\nonumber\\
&\quad\frac{\Omega _s^2}{\gamma _{32}+i \left(\Delta _c-\Delta _{p2}-\Delta _s\right)}+\frac{\Omega _s^2 \left(\left(i \gamma _{43}-\Delta _c+\Delta _{p2}\right) \left(-i \gamma _{32}+\Delta _c-\Delta _{p2}-\Delta _s\right)-\Omega _c^2+\Omega _s^2\right)}{\left(\gamma _{32}+i \left(\Delta _c-\Delta _{p2}-\Delta _s\right)\right) \left(\left(-i \gamma _{32}+\Delta _c-\Delta _{p2}-\Delta _s\right) \left(i \gamma _{52}+\Delta _{p2}+\Delta _s\right)+\Omega _c^2-\Omega _s^2\right)}\biggr)\\
\noalign{\medskip}
B&=\left(\gamma _{32}+\gamma _{54}+i \left(\Delta _c-2 \Delta _{p2}-\Delta _s\right)\right) \Omega _s^2 \left(\left(i \gamma _{43}-\Delta _c+\Delta _{p2}\right) \left(-i \gamma _{32}+\Delta _c-\Delta _{p2}-\Delta _s\right)-\Omega _c^2+\Omega _s^2\right)+\nonumber\\
&\quad\left(\left(-i \gamma _{32}+\Delta _c-\Delta _{p2}-\Delta _s\right) \left(i \gamma _{52}+\Delta _{p2}+\Delta _s\right)+\Omega _c^2-\Omega _s^2\right) \bigl(\left(\gamma _{32}+i \left(\Delta _c-\Delta _{p2}-\Delta _s\right)\right)\times \nonumber\\
&\quad\left(\left(-i \gamma _{43}+\Delta _c-\Delta _{p2}\right) \left(i \gamma _{54}+\Delta _{p2}\right)+\Omega _c^2\right)+\left(\gamma _{54}-i \Delta _{p2}\right) \Omega _s^2\bigr)
\end{align}
\end{subequations}
\end{widetext}



\end{document}